\documentclass[12pt, amsmath, amssymb]{iopart}

\usepackage[dvips]{graphicx}
\usepackage{mathrsfs}
\usepackage{graphicx}
\usepackage{amssymb}
\usepackage{epsf,latexsym}
\usepackage{times}
\usepackage{ifpdf}
\ifpdf
\usepackage{epstopdf}   
\fi

\newcommand{\bra}[1]{\langle#1|}
\newcommand{\ket}[1]{|#1\rangle}

\newcommand{\sz}[0]{\ensuremath{\mathbf{\sigma}_z}}

\newcommand{\be}{\begin{equation}}
\newcommand{\ee}{\end{equation}}
\newcommand{\ba}{\begin{eqnarray}}
\newcommand{\ea}{\end{eqnarray}}

\usepackage{color}

\begin{document}

\bibliographystyle{unsrt}

\title{Aspects of Quantum Coherence in Nanosystems}

\author{Brendon W. Lovett}
\ead{brendonlovett@materials.ox.ac.uk}
\address{Department of Materials, University of Oxford, Parks Road, Oxford, OX1 3PH, United Kingdom}

\author{Ahsan Nazir}
\ead{ahsan.nazir@ucl.ac.uk}
\address{Department of Physics and Astronomy, University College London, Gower Street, London WC1E 6BT, United Kingdom}

\date{\today}

\begin{abstract}

Coherence is a familiar concept in physics: It is the driving force behind wavelike phenomena such as the diffraction of light. Moreover, wave-particle duality implies that all quantum objects can exhibit coherence, and this {\it quantum coherence} is crucial to understanding the behaviour of a plethora of systems. 

In this article, which is written at an undergraduate level, we shall briefly introduce what is meant by coherence in a well-known classical setting, before going on to describe its quantum version. We will show that coherence is important in describing the properties of solid-state nanosystems, and especially quantum dots. Simple experiments that reveal the coherent nature of matter  -- and how this leads to some very powerful applications -- will be described. Finally, we shall discuss the fragility of coherence and shall introduce a method for describing decoherence in open quantum systems.

\end{abstract}

\maketitle

\section{Introduction}

Nanosystems, comprising perhaps a few thousand atoms, are by their very nature small. So small, in fact, that quantum mechanics often plays a fundamental role in determining their properties and behaviour. For example, quantum confinement in nanostructures can lead to a discrete energy spectrum, much like in atoms. A particularly exciting consequence of this discretization is the possibility of creating and observing quantum coherent superpositions of these energy states~\cite{Borri07}, which could then be further manipulated in order to perform computational tasks; this is termed quantum information processing (QIP)~\cite{Nielsen00}). Coherence is key in such ambitious applications, but it must be maintained on relatively long timescales to give a reasonable chance of any task being realised. This is easier said than done. As any quantum system is enclosed by an environment, this naturally leads to a leakage of coherence from the nanosystem into its surroundings, in a process known as decoherence. Once decoherence has taken its toll, the system behaviour becomes essentially classical~\cite{Zurek91}, and useless as far as quantum information applications are concerned.

In this article, we will provide a concise introduction to the ideas of coherence and decoherence in nanosystems, with an emphasis on their conflicting roles in determining the feasibility of performing QIP tasks in such a setting. We start, in Section~\ref{definition}, with a definition of coherence in terms of the classical physics of waves, before moving into the quantum regime. In Section~\ref{coherencenano} we discuss some basic aspects of coherence in nanostructures, while in Section~\ref{applications} we introduce a simple application. Finally, in Section~\ref{decoherence}, we explore the destructive power of decoherence processes and outline one of the main techniques used in modeling them.

\section{Definition of coherence}\label{definition}

The first encounter we normally have with the phenomenon of coherence in physics is in studying wave motion. A wave can be regarded as coherent if its value at one point in space and time can be related precisely to its value at nearby points. The larger the spatial and temporal extent of this relation, the more coherent the wave is. For a sine wave, we can be quite specific. The argument of the sine function, called the {\it phase} of the wave, should be a well defined function of space and time if the wave is to be coherent.

\begin{figure}
\begin{center}
\includegraphics[width=0.8\columnwidth]{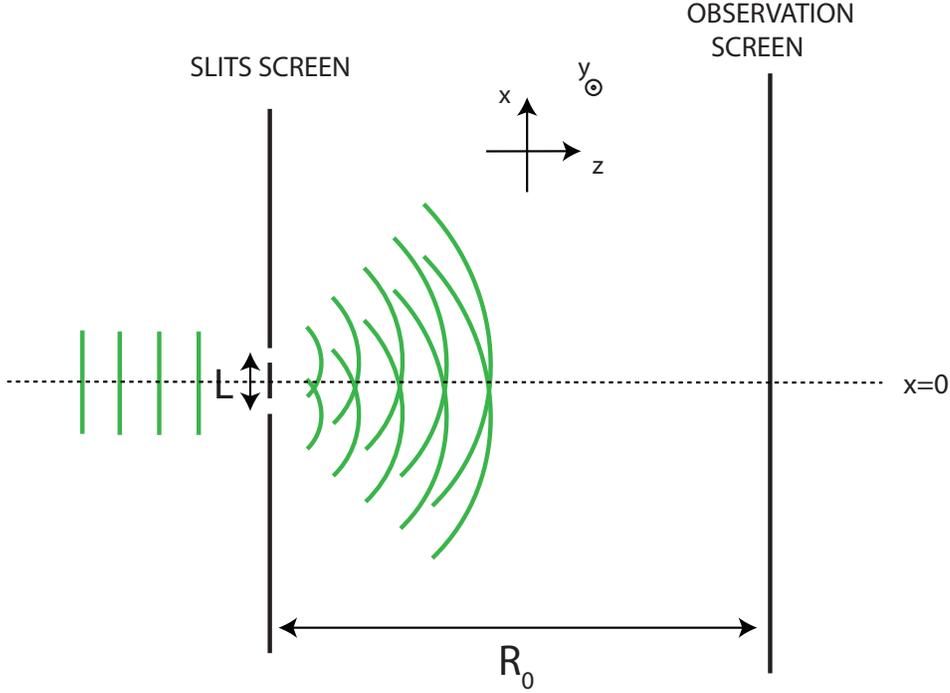}
\caption{Schematic diagram of the experimental set-up needed for the observation of the interference pattern from Young's double slits. The interference fringes are still visible, even when photons pass through the apparatus one at a time, and this is a consequence of quantum coherence.} \label{fig:slits}
\end{center}
\end{figure}

Coherence is particularly important when we consider how two different waves {\it interfere} with one another. In fact, for interesting effects to be observed, the waves must be both self and mutually coherent. The simplest example of this is in the famous Young's double slit experiment. The two light rays emerging from each slit can each be described by an oscillating electric field with wave vectors ${\bf k}$ and ${\bf k}'$. Typically both fields originate from the same source and so their wave vectors have equal magnitude $k$ and they have equal optical frequencies $\omega$. We may write the two fields at point ${\bf r}$ and time $t$ as
\ba
{\bf E}_1 = E_0 {\bf j} \exp(-i ({\bf k}.{\bf r} + \omega t)), \nonumber \\
{\bf E}_2 = E_0 {\bf j} \exp(-i ({\bf k}'.{\bf r} + \omega t)).
\ea
Where we have also assumed that the experiment is set up so that the field amplitude in both rays is equal, and that the field is polarized parallel with the slits along the $y$ axis (labelled ${\bf j}$). If we now wish to calculate the intensity pattern at an observation screen at a position ${\bf R}_0$ relative to the slit screen, we can simply add the two fields. At point $x {\bf i}$ along the axis of the screen (see Fig.~\ref{fig:slits}) we have
\be
{\bf E}_T = E_0 {\bf j} \exp(-i\omega t + k_z R_0) (1 + \exp(-i \Delta k_x x)), 
\ee
where $k_z$ is the $z$ component of the wave vector (which is the same for both beams if we assume the screen is sufficiently far from the slits) and $\Delta k_x = k_x - k_x'$ is the difference of the two wavevector components in the $x$ direction. The quantity $\Delta k_x x$ represents the {\it phase difference} of the two beams, and we shall see that it is crucial in determining how the two waves interfere.

The intensity $I$ of the pattern on the screen is proportional to the square modulus of the field and so we can write it as:
\be
I\propto | 1 + \exp(-i \Delta k_x x)|^2 = 2 \cos^2(\Delta k_x x) = 2\cos^2 \left( \frac{k L x}{R_0}\right)
\ee
with $L$ the slit spacing. We have assumed that $L\ll R_0$. The intensity is maximum when the two waves constructively interfere with one another - i.e. when they have the same phase up to a multiple of 2$\pi$. We see therefore that in this example the phase difference between the two waves completely determines the interference pattern.

So far, we have been able to describe everything using classical fields. However, in this article we wish to introduce the concept of {\it quantum coherence} and we shall now see how it arises.

It is well known that Young's interference pattern is still seen even when the flux of light through the slits is so small that there is, on average, well below one light quantum, or {\it photon}, in the system at any given time. In this case, it is not possible to think of the interference as arising from two distinct fields, which can have varying amplitude. We must instead turn to the concept of quantum superposition -- the quantum property that allows a particle to go through both slits, and then `interfere with itself' on the screen. Put more formally, the particle is said to be in a linear superposition of eigenstates of the measurement operator that would determine whether the particle passes through one slit or the other - or the `which slit?' operator.

Let us label a photon which passes through slit one as $\ket{\bf k}$, and that which passes through slit two as $\ket{{\bf k}'}$. In second quantized notation, these states can also be written in terms of creation operators acting on the vacuum, for example $\hat{a}^\dagger_{\bf k} \ket{0}=\ket{{\bf k}}$, where the operator  $\hat{a}^\dagger_{\bf k}$ is said to create a photon in the mode labelled ${\bf k}$.

Quantum superposition allows us to also write a single photon as a linear combination of these states:
\be
\ket{\psi} = a \ket{{\bf k}} + b \ket{{\bf k}'}.
\label{state}
\ee
where normalization means that  $a^2 + b^2 = 1$. In general $a$ and $b$ are complex numbers -- but for simplicity we will assume they are real for now, introducing the complications of complex representations later.

In order to calculate the intensity of the screen pattern, we must use the electric field operator which, in terms of our creation operator is $\hat E \propto \sum_{\bf k} a_{\bf k} e^{-i(\omega_{\bf k} t + {\bf k}\cdot {\bf r})} + a_{\bf k}^\dagger e^{i(\omega_{\bf k} t  -{\bf k}\cdot {\bf r})}$~\cite{walls08}. The intensity profile is found, then, by taking the expectation value of the square of the field operator~\cite{walls08}, for the state $\ket{\psi}$. After removing some terms which must be zero, we find
\be
I \propto \bra{\psi} a_{\bf k'}^\dagger a_{\bf k} e^{i({\bf k'-k})\cdot{\bf r}} + a_{\bf k}^\dagger a_{\bf k'} e^{i({\bf k-k'})\cdot{\bf r}}  + a_{\bf k'}^\dagger a_{\bf k'} + a_{\bf k}^\dagger a_{\bf k} \ket{\psi}  .
\label{qintense}
\ee
Using our state, Eq.~\ref{state}, in Eq.~\ref{qintense}, we obtain
\be
I \propto 1 + 2ab \cos(({\bf k} - {\bf k'})\cdot{\bf r}).
\ee
This tells us that if the two components of the superposition are equal ($a=b=1/\sqrt{2}$), then we reproduce the classical wave pattern we found earlier. If, on the other hand, the system is an eigenstate of the `which slit?' operator, then we lose the pattern altogether, since either $a$ or $b$ are zero. This is exactly what we know to be true: determining through which of the two slits the photon passed destroys the interference effect. We therefore see that in order to observe effects due to  quantum coherence we must have a superposition. Further, the relative phase of the components of the superposition is key to calculating just what the pattern is. In our example, adding an extra phase factor and so making one of the coefficients complex:
\be
\ket{\psi} = a \ket{{\bf k}} + be^{i\phi} \ket{{\bf k}'},
\label{state2}
\ee
results in a shifting of the pattern:
\be
I \propto 1 + 2ab \cos(({\bf k} - {\bf k'})\cdot{\bf r}-\phi).
\ee

Quantum coherence is therefore not just a convenient mathematical tool for describing solutions to Schr\"odinger's equation: It actually gives rise to distinct, observable experimental features. We shall see in the next section that it is important for describing a broad class of physical phenomena.

\section{Coherence in nanostructures}\label{coherencenano}

All we have encountered so far is familiar from high school physics, even if we have set the ideas on a more mathematical foundation. However, the effect of quantum coherence is not limited to optics experiments, which clearly have a wave-like character. In fact, quantum coherence is vitally important in understanding the electronic properties of solid state structures, and in particular solid state nanostructures such as {\it quantum dots} (QDs).

\begin{figure}
\begin{center}
\includegraphics[width=0.95\columnwidth]{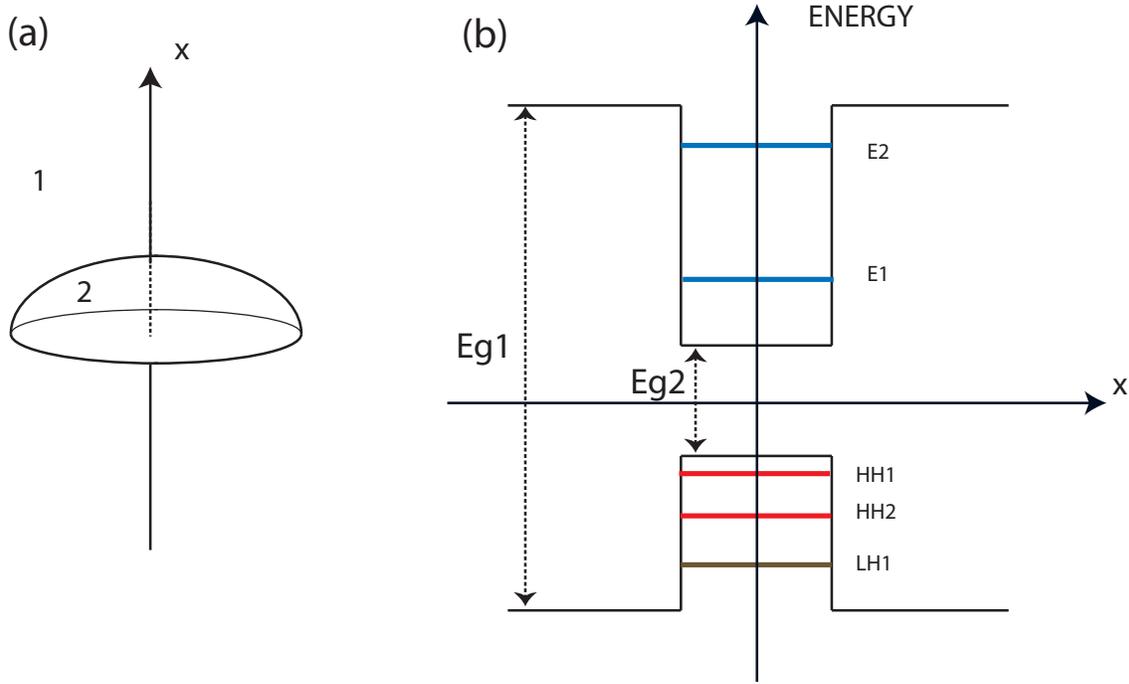}
\caption{(a) A QD is a nanoscale region (labelled 2 here) of a smaller bandgap semiconductor that is encased in a larger bandgap semiconductor (labelled 1 here). They come in many different shapes, the dome shape depicted here being one of the more common. (b) The profile of band edges along a cut through the axis of the QD. The two bandgaps are labelled $Eg1$ and $Eg2$. The QD has confined states in both the conduction band (here labelled $E$) and valence band (here labelled $LH$ and $HH$).} \label{fig:semicon}
\end{center}
\end{figure}

QDs are usually made of semiconductor materials, which as we know have a forbidden energy gap (the band gap) between the highest lying valence states and lowest lying conduction states. The particular type of QD that we will consider is a nano-sized region of a semiconductor, typically made of 1000 to 10000 atoms, that is encapsulated in another type of semiconductor~\cite{jacak98,skolnick04}. Such structures are {\it self-assembled} since they form naturally under certain growth conditons. The surrounding material generally has a much larger band gap than that in the QD. This means that forbidden states in the surrounding material are not forbidden in the QD, and effectively the QD becomes a potential well for electrons in both the conduction and valence bands (see Fig.~\ref{fig:semicon}): This results in {\it confinement} of the particles. 

The energy states of the confined states are often well described by a simple undergraduate level solution to the Schr\"odinger equation for a finite potential well. This, of course, gives us discrete energy levels, both above and below the energy gap of the QD (again, see Fig.~\ref{fig:semicon}). These discrete, confined levels have much in common with those in an atom -- and so a QD is often termed an `artificial atom'; indeed we may use atomic models to describe QDs. 

We wish to think about coherence, which as we know requires superposition, so the simplest model we can consider is one which has two levels. The most natural choice would be the ground state $\ket{g}$ of the QD -- the one in which all of the states below the band gap are full -- and the first excited state $\ket{e}$. $\ket{e}$ would have a single electron in the state immediately above the band gap, and would leave behind an unfilled state (or {\it hole}) in the state just below the band gap. Such an electron-hole pair is known as an exciton and is typically an electron-volt or two above $\ket{g}$. This energy corresponds to the optical range of the electromagnetic spectrum.

\begin{figure}
\begin{center}
\includegraphics[width=0.7\columnwidth]{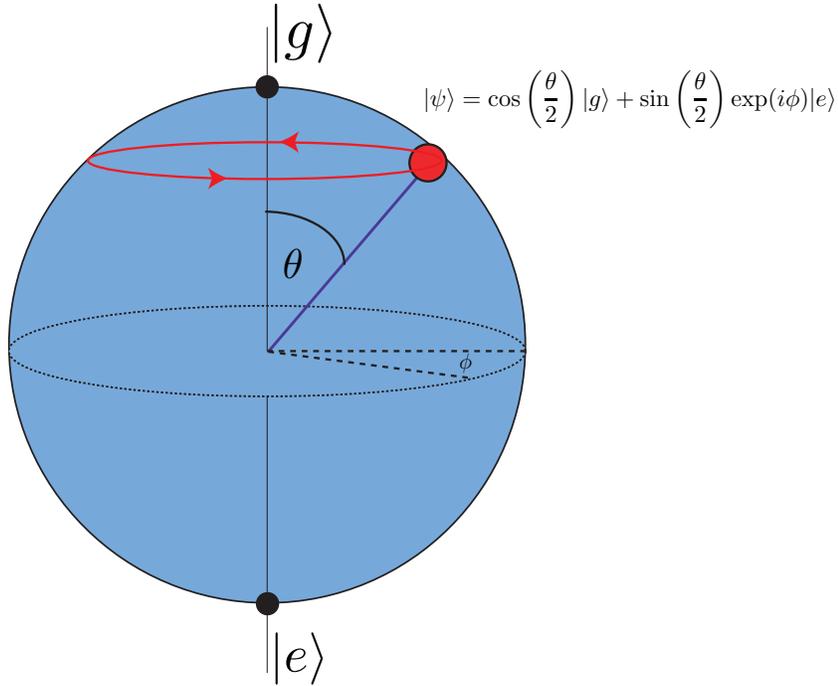}
\caption{The {\it Bloch sphere} is a representation of a two level system. The states $\ket{g}$ and $\ket{e}$ are at the north and south poles and superposition states are characterized by the polar angle $\theta$ and azimuthal angle $\phi$. The time evolution of a general state $\ket{\psi}$, depicted as a red curve, is a rotation about the axis connecting the eigenstates -- in this case the vertical axis connecting $\ket{g}$ and $\ket{e}$.} \label{bloch}
\end{center}
\end{figure}

As we know, the general state of this two level system can be represented by the equation
\be
\ket{\psi} = a\ket{g} + b\ket{e},
\label{psicomplex}
\ee
where we now allow $a$ and $b$ to be complex (so $|a|^2+|b|^2 = 1$). The state of this system can also be written in terms of two angles $\theta$ and $\phi$:
\be
\ket{\psi} = \cos\left(\frac{\theta}{2}\right) \ket{g} + \sin\left(\frac{\theta}{2}\right)\exp(i\phi) \ket{e},
\ee
which then automatically satisfies the normalization condition and also allows us the freedom to represent different phases between the two components in the superposition. We are now able to represent our state on the surface of a sphere - the {\it Bloch sphere} with polar angle $\theta$ and azimuthal angle $\phi$. $\ket{g}$ is then the `north pole' and $\ket{e}$ the `south pole' (see Fig.~\ref{bloch}).

Where is the quantum coherence in this picture? First consider a measurement that determines whether the system is in state $\ket{0}$ or $\ket{1}$; it gives the result $\ket{0}$ with probability $\cos^2(\theta/2)$ and $\ket{1}$ with probability $\sin^2(\theta/2)$. Such statistics can straightforwardly be generated by a classical system, and so are not inherently quantum. On the other hand, the angle $\phi$, the measure of the phase difference between $\ket{e}$ and $\ket{g}$, is a uniquely quantum property of our particle, and as we shall see many uniquely quantum effects rely on it having a well defined value.

Let us now discuss an experiment whose results depend on $\phi$. As we know, $\ket{g}$ and $\ket{e}$ are eigenstates of our QD square well Hamiltonian. The time dependent Schr\"odinger equation therefore tells us that they are eigenstates of the evolution operator $i \hbar \partial /\partial t$ with an eigenvalue corresponding to the energy of the state. If we start out in a state with $\theta= \theta_0$ and $\phi = 0$ the subsequent time evolution is
\be
\ket{\psi} = \cos\left(\frac{\theta_0}{2}\right) \ket{g} + \sin\left(\frac{\theta_0}{2}\right)\exp\left(\frac{i\Delta t}{\hbar}\right) \ket{e},
\ee
with $\Delta$ being the energy difference between the states. We have dropped an overall `global' phase factor since only relative phase factors have physical meaning. We see that the time evolution corresponds to a rotation of the state on the Bloch sphere, around the axis connecting the two eigenstates, with angular frequency $\Delta/\hbar$ -- and a measurement of an oscillation at this frequency is a signature of quantum coherence.

It tends to be easiest to make measurements in the eigenbasis -- i.e. to say whether our system is in $\ket{g}$ or $\ket{e}$. Since such a measurement does not depend on the phase, a clever trick has to be used to detect it. A full description of the experiment is beyond the scope of this paper, but the crucial element is that an optical frequency laser, tuned to the energy difference between $\ket{g}$ and $\ket{e}$, makes the system oscillate between the two states. It is equivalent to a rotation around the $y$ axis of our Bloch sphere. Now consider the following experiment. First, our system is cooled to the ground state. Second, a laser pulse is applied long enough to produce an equal superposition of $\ket{g}$ and $\ket{e}$. Third, the system is allowed to evolve for a time $\tau$ under its own steam, and therefore rotates along the equator by an angle $\Delta \tau/\hbar$. Next, a second pulse is applied that is the same as the first; if the state has made a full revolution, we end up in $\ket{e}$, if only half a revolution has occured the pulse takes us back to $\ket{g}$. We next readout the state, and after repeating the experiment many times we will see oscillations in the populations of $\ket{g}$ and $\ket{e}$ as a function of the time delay $\tau$. These oscillations, also known as `Ramsey fringes', have been measured (see for example Stufler {\it et al.}~\cite{stufler06}) and are direct evidence for quantum coherence.

\section{Applications of coherence and quantum computing}\label{applications}

What is coherence good for? In recent years, the field of QIP has led to some interesting possibilities. In this section, we'll focus on one example: quantum superdense coding~\cite{bennett92b}.

Let us imagine we have two people, called Alice $A$ and Bob $B$ who wish to communicate a message to each other. Each of them possesses a quantum two-level system (also called a {\it qubit}) as described in the previous section. (In fact, a more mobile qubit would be more suitable for this application, but it illustrates the power of coherence; we shall discuss applications of nanostructures shortly). 

Since we are now moving into a discussion of information, we shall switch our labels from $\{g,e\}$ to $\{0,1\}$. The general joint state of the $A$ and $B$ qubits can then be represented as a superposition of four basis states:
\be
\ket{\psi} = a\ket{00} + b\ket{01} + c\ket{10} + d\ket{11}.
\ee
where the first label for each component represents $A$ and the second $B$. Coefficients $a, b, c$ and $d$ can of course be complex as each part of the superposition can have a different relative phase. Now consider the state $(\ket{00}+\ket{11})/\sqrt{2}$. This kind of state cannot be represented by a product of a state of qubit A and a state of qubit B, and for this reason is called {\it entangled}. If $A$ and $B$ are measured the outcomes are highly correlated with one another -- a further consequence of the coherence between the states.

Now, by simply applying different {\it local} operations to her qubit, Alice can choose to make, for example, the following four states, which she can associate with four different bit strings:
\ba
\frac{1}{\sqrt{2}}(\ket{00}+\ket{11}) &\,\,\,\,\,\,\,\,\,\,& (00);\nonumber\\
\frac{1}{\sqrt{2}}(\ket{00}-\ket{11}) & &(01);\nonumber\\
\frac{1}{\sqrt{2}}(\ket{01}+\ket{10}) & &(10); \nonumber\\
\frac{1}{\sqrt{2}}(\ket{01}-\ket{10}) & &(11) . \label{entangledstates}
\ea
Now, having performed her manipulation, Alice passes on her qubit to Bob. Bob can then measure both qubits together and, since the four states above are orthogonal to one another, can distinguish between them. We therefore conclude that, with the transfer of only a single quantum bit, Alice can communicate two classical bits worth of information. The protocol only works because Alice and Bob's qubits share quantum coherence before the protocol is carried out.

\begin{figure}
\begin{center}
\includegraphics[width=0.5\columnwidth]{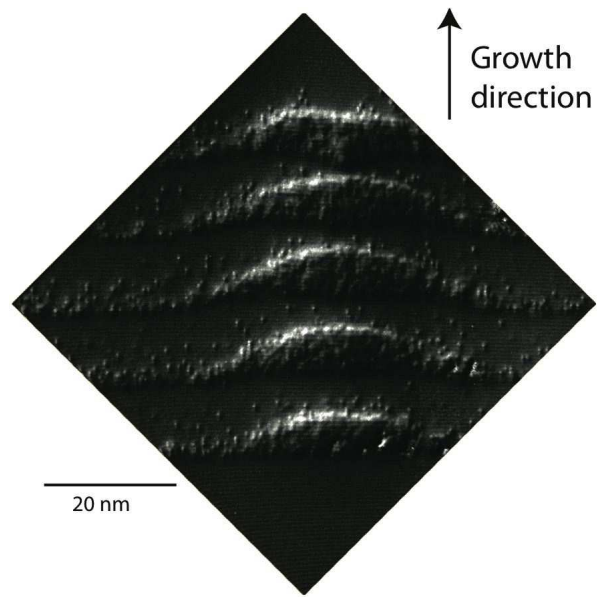}
\caption{A cross-section scanning tunnelling micrograph of a stack of InAs dots in a GaAs matrix. Figure reused with permission from D. M. Bruls, Applied Physics Letters, 82, 3758 (2003). Copyright 2003, American Institute of Physics~\cite{bruls03}.}
 \label{dots}
\end{center}
\end{figure}

This simple example illustrates how careful manipulation of coherence can lead to enhanced processing capabilities. In fact, there exist much wider possibilities. A major goal of nanotechnological research is to build structures that can be made to house a whole array of qubits, each of which can be individually manipulated and made to interact controllably with its neighbours. Such a structure would be a basic quantum processor, upon which quantum algorithms can be performed. An example of one prototype structure is shown in Fig.~\ref{dots}, which shows a real electron micrograph of a stack of QDs.

Quantum algorithms with significant speed-up over classical versions have been proposed for a number of tasks~\cite{galindo02}. They seem to be particularly suited to tasks for finding information, where a large superposition can be used to embody a vast amount of information - and then quantum coherence is used to `bring out' certain states. Prime examples are the factorization of large numbers~\cite{shor94}, and algorithmic searching~\cite{grover97}.

\section{Mixed states and decoherence}\label{decoherence}

So far, we have been interested in the properties and applications of quantum coherence in the idealised case of simple, isolated, two-level systems representing qubits. Such closed systems may be completely described by a pure state $|\psi\rangle$, or equivalently a wavefunction $\psi(t)$, and their evolution is governed by the Schr\"odinger equation. However, in practice, it is never truly possible to isolate a particular quantum system from its surroundings, and we must think instead in terms of an {\it open} quantum system describing our qubit interacting with its environment~\cite{breuer02}. For example, in a semiconductor nanosystem such as a QD, the excitonic states that embody our qubit may interact with the vibrational (phonon) states of the underlying semiconductor solid-state lattice. If these interactions are significant (which they generally are, even at extremely low temperatures~\cite{Borri07}), then it is clear that we are no longer valid in considering our two-level qubit as an isolated, closed system. 

The coupling of qubit and environment has the potential to cause a transfer of quantum coherence from the system of interest into its surroundings. As we typically have no way of keeping track of the environmental state, we must consider this coherence to be lost from our qubit (mathematically, this is represented by taking a trace over the environmental degrees of freedom). This process is known as decoherence, and is currently one of the greatest obstacles to large-scale QIP in the solid-state. As an example, the excitonic states of self-assembled QDs in Gallium Arsenide typically have decoherence times $\sim 1$~ps - $1$~ns~\cite{Borri07}, meaning that their quantum coherence is suppressed exponentially on these timescales. This obviously leaves very little time with which to exploit the system's coherent properties, and hence the detailed understanding of decoherence processes, with the aim of then subverting them, has become a problem of considerable interest to the research community over the past few years.  

In order to illustrate the effects of decoherence in open quantum systems it is useful to introduce a statistical representation of quantum states, known as the density operator.  As we will shortly see, the density operator is particularly useful in describing states about which we do not have complete knowledge. We begin, however, by considering a familiar pure state, and define the density operator simply as an alternative description of the system through
\begin{equation}\label{rhopure}
\rho=|\psi\rangle\langle\psi |,
\end{equation}
which for the two-level example of Eq.~\ref{psicomplex} becomes
\begin{equation}
\rho=|a|^2|g\rangle\langle g|+ab^*|g\rangle\langle e|+a^*b|e\rangle\langle g|+|b|^2|e\rangle\langle e|.
\end{equation}
Using the vector representations of $|g\rangle$ and $|e\rangle$:
\begin{equation}\label{gevecs}
|g\rangle=\left(
  \begin{array}{ c }
     1  \\
     0 
  \end{array} \right),
  |e\rangle=\left(
  \begin{array}{ c }
     0  \\
     1 
  \end{array} \right),
\end{equation}
we may re-write the density operator as a matrix (with elements $\rho_{00}$, $\rho_{01}$, $\rho_{10}$, and $\rho_{11}$)
\begin{equation}\label{rhomat}
\rho=\left(\begin{array}{ c }
     a  \\
     b 
  \end{array} \right)\left(\begin{array}{ c c }
     a^* & b^* 
  \end{array} \right)=\left(\begin{array}{ c c }
     |a|^2 & ab^*\\
     a^*b & |b|^2
  \end{array} \right)=\left(\begin{array}{ c c }
     \rho_{00} & \rho_{01}\\
    \rho_{10} & \rho_{11}
  \end{array} \right).
\end{equation}
Here, the diagonal elements $|a|^2$ and $|b|^2$ are simply the probabilities that a measurement of the system state will return $|g\rangle$ or $|e\rangle$ respectively, and are known as \emph{populations}. The off-diagonal elements, meanwhile, describe the phase relationship between $|g\rangle$ and $|e\rangle$ in the state $|\psi\rangle$, and are therefore termed \emph{coherences}.

Imagine now that our system qubit is not isolated, but is actually coupled to an external environment. As a particularly simple example, we could consider the environment to be comprised of a single qubit as well. In this case, interactions between the system qubit ($s$) and environment qubit ($\xi$) can lead to entangled system-environment states, such as those encountered in Eq.~\ref{entangledstates}. By way of illustration, let's assume that qubit $\xi$ takes state $|0\rangle$ for system state $|g\rangle$, and $|1\rangle$ for system state $|e\rangle$, in which case the joint system-environment state may be written
\begin{equation}\label{sys-envsinglequbit}
|\Psi\rangle=c_{g0}|g\rangle_s\otimes|0\rangle_{\xi}+c_{e1}|e\rangle_s\otimes|1\rangle_{\xi}=c_{g0}|g0\rangle+c_{e1}|e1\rangle,
\end{equation}
where $|c_{g0}|^2+|c_{g1}|^2=1$. This leads to a density operator
\begin{eqnarray}\label{sys-envsinglequbitrho}
\rho&{}={}&|\Psi\rangle\langle\Psi|=|c_{g0}|^2|g0\rangle\langle g0|+c_{g0}c_{e1}^*|g0\rangle\langle e1|+c_{g0}^*c_{e1}|e1\rangle\langle g0|+|c_{e1}|^2|e1\rangle\langle e1|\nonumber\\
&{}={}&\left(\begin{array}{ c c c c}
     |c_{g0}|^2 & 0 & 0 & c_{g0}c_{e1}^* \\
     0 & 0 & 0 & 0 \\ 
     0 & 0 & 0 & 0 \\ 
     c_{g0}^*c_{e1} & 0 & 0 & |c_{e1}|^2
  \end{array} \right),
\end{eqnarray}
where, once again, the diagonal elements are populations and the off-diagonal coherences.
As mentioned earlier, it seems plausible to believe that we have no way of observing the environmental state and can make measurements only on the system qubit $s$. In this situation, we express our ignorance of the state of $\xi$ by constructing a density operator for the system qubit alone, that completely represents outcomes of all possible measurements local to that qubit. The prescriptive procedure for this is called the partial trace, and on taking such a trace of $\rho$ over the environmental degrees of freedom (here, the states $|0\rangle_{\xi}$ and $|1\rangle_{\xi}$) we find a \emph{reduced} system density operator
\begin{equation}\label{reducedrho}
\rho_s={\rm tr}_{\xi}\rho=|c_{g0}|^2|g\rangle\langle g|+|c_{e1}|^2|e\rangle\langle e|
=\left(\begin{array}{ c c}
     |c_{g0}|^2 & 0 \\
     0 & |c_{e1}|^2
  \end{array} \right).
\end{equation}
We see that the orthogonality of the environment states $|0\rangle_{\xi}$ and $|1\rangle_{\xi}$ has led to a complete destruction of the system coherences in the reduced density operator. The state $\rho_s$ defined in Eq.~\ref{reducedrho} is therefore an example of a \emph{mixed} state as it cannot be written as a pure state vector $|\psi\rangle$. It expresses the fact that, in this example, measurements of the system state alone can give us only the classical outcomes $|g\rangle$ or $|e\rangle$ with probabilities $|c_{g0}|^2$ and $|c_{e1}|^2$, respectively. Remarkably, though the combined state of Eq.~\ref{sys-envsinglequbitrho} is pure, our inability to observe the environmental state condemns us to be unable to infer any system coherence whatsoever.

The complete loss of coherence in the above example is particularly severe. However, we may generalise the calculation slightly and instead of considering an environment consisting of a single qubit, treat one comprising of many basis states:\begin{equation}\label{sys-envmany}
|\Psi\rangle=c_{g0}|g\rangle_s\otimes|\xi_0\rangle+c_{e1}|e\rangle_s\otimes|\xi_1\rangle=c_{g0}|g\xi_0\rangle+c_{e1}|e\xi_1\rangle,
\end{equation}
where $\xi_0$ and $\xi_1$ symbolically represent different environmental states corresponding to system states $|g\rangle$ and $|e\rangle$, respectively. It can be shown that the reduced system density operator in this case is given by~\cite{breuer02}
\begin{equation}\label{reducedrhomany}
\rho_s={\rm tr}_{\xi}\rho
=\left(\begin{array}{ c c}
     |c_{g0}|^2 & c_{g0}c_{e1}^*\langle\xi_1|\xi_0\rangle \\
     c_{g0}^*c_{e1}\langle\xi_0|\xi_1\rangle & |c_{e1}|^2
  \end{array} \right),
\end{equation}
where the loss of system coherence is now determined by the overlap of the environmental states through $\langle\xi_0|\xi_1\rangle$. For most physical systems, this overlap  will steadily decrease in time leading to a continual suppression of the system coherences, known as dephasing or decoherence. In fact, the example studied here is usually termed \emph{pure} dephasing since the environmental processes induce no changes in the system populations, only in the coherences. This type of decoherence is particularly important in semiconductor QDs, where it is thought to arise through exciton-phonon interactions and can dominate dephasing of the dot states at short times and low temperatures~\cite{Borri07}.

The question then naturally arises of how might we model the dynamics of our qubit subject to decoherence? The detailed answer depends on the system under consideration and unfortunately lies well beyond the scope of this article (see Ref.~\cite{chirolli08} for a thorough recent review). However, we may introduce one important technique, that of \emph{master equations}~\cite{breuer02,Gardiner04,Carmichael93}, on a much more phenomenological level. The density operator of a closed quantum system evolves according to the Liouville-von Neumann equation~\cite{breuer02}
\begin{equation}\label{von-neumann}
\frac{d\rho}{dt}=-\frac{i}{\hbar}[H,\rho],
\end{equation}
where $H$ is the system Hamiltonian. It turns out that for open quantum systems an accurate description
of the reduced system dynamics under the influence of the external environment can often be given by a master equation of the form
\begin{equation}\label{lindbladmaster}
\frac{d\rho_s}{dt}=-\frac{i}{\hbar}[H,\rho_s]+\sum_k\left(L_k\rho_sL_k^{\dagger}-\frac{1}{2}(L_k^{\dagger}L_k\rho_s+\rho_sL_k^{\dagger}L_k)\right).
\end{equation}
Here, $L_k$ are termed Lindblad operators and account for the impact of the system-environment coupling on the otherwise closed system described by $H$. 

The number and form of Lindblad operators necessary to model a particular physical situation depends on the precise nature of the system-environment interactions involved.
As an example, let's consider the decoherence process represented by a single Lindblad operator of form
\begin{equation}\label{lindblad}
L=\sqrt{\delta}\sz,
\end{equation} 
where
\begin{equation}\label{sz}
\sigma_z=\left(\begin{array}{ c c}
    1 & 0 \\
     0 & -1
  \end{array} \right)
\end{equation}
is one of the Pauli matrices. How might such a Lindblad operator arise? Well, the Hamiltonian of an isolated two-level system can be given as $H=(\epsilon/2)\sigma_z$, where $\epsilon$ is the energy difference between the ground and excited states, $|g\rangle$ and $|e\rangle$, which themselves are eigenvectors of $H$. Hence, a Lindblad operator of the form $\sqrt{\delta}\sz$ naturally represents stochastic fluctuations in the energy separation between $|g\rangle$ and $|e\rangle$, with an associated rate $\delta$, and so leads to uncertainty in the phase evolution of the system state. We might therefore expect that such a Lindblad operator should lead to a description of pure dephasing, and we will now show that this is indeed the case. Inserting the above forms of $L$ and $H$ into Eq.~\ref{lindbladmaster} we find
\begin{equation}\label{puredephasemaster}
\frac{d}{dt}\left(\begin{array}{ c c }
     \rho_{00} & \rho_{01}\\
    \rho_{10} & \rho_{11}
  \end{array} \right)=\left(\begin{array}{ c c }
    0 & -(2\delta+i\epsilon/\hbar)\rho_{01}\\
    -(2\delta-i\epsilon/\hbar)\rho_{10} & 0
  \end{array} \right),
\end{equation}
which gives the solutions
\begin{eqnarray}\label{puredephasesols}
\rho_{00}(t)&{}={}&\rho_{00}(0),\nonumber\\
\rho_{01}(t)&{}={}&e^{-2\delta t}e^{-i\epsilon t/\hbar}\rho_{01}(0),\nonumber\\
\rho_{10}(t)&{}={}&e^{-2\delta t}e^{i\epsilon t/\hbar}\rho_{10}(0),\nonumber\\
\rho_{11}(t)&{}={}&\rho_{11}(0),
\end{eqnarray}
where
\begin{equation}
\rho(0)=\left(\begin{array}{ c c }
     \rho_{00}(0) & \rho_{01}(0)\\
    \rho_{10}(0) & \rho_{11}(0)
  \end{array} \right)
\end{equation}
is the initial system density operator.
We see that, as anticipated, a Lindblad operator proportional to $\sigma_z$ leads to a model where the system populations do not change in time, while the coherences are suppressed at a rate $2\delta$, i.e. to a description of the dynamics of pure dephasing, with a characteristic time $T_2=1/2\delta$.

Finally, it is worth noting that by including a suitable choice of control in the system Hamiltonian, it is possible to use the master equation to assess the performance of complete quantum gates subject to decoherence. For example, external laser control of QD excitonic states can be represented by the addition of a term of the form $(\Omega\cos{\omega_0 t})\sigma_x$ in $H$, where
\begin{equation}\label{sx}
\sigma_x=\left(\begin{array}{ c c}
    0 & 1 \\
     1 & 0
  \end{array} \right)
\end{equation}
is another of the Pauli matrices. Here, $\omega_0$ is the laser frequency and $\Omega$ is the laser-dot coupling strength, known as the Rabi frequency. Such a term naturally leads to coherent oscillations in the system populations (i.e. to single qubit gates) on a timescale $\sim 1/\Omega$, that become damped over time due to the dephasing process. The smallness of the ratio of gate operation time to decoherence time is often used as a figure of merit to assess the feasibility of performing quantum operations with a particular setup, and here would be of order $\delta/\Omega$. We see then that for a system to be a good candidate quantum gate, it must be controllable on a much shorter timescale than 
that set by the dominant decoherence processes. In self-assembled QDs, picosecond control of the excitoinc states is feasible, with decoherence times approaching nanoseconds at very low temperature~\cite{Borri07}. This gives a potential figure of merit $\sim 10^{-3}$, which is sufficient for many quantum gates to be performed. Perhaps more promising, however, is the proposed use of confined QD electron or hole spin degrees of freedom as qubit states, where coherence times of the order of microseconds or greater may perhaps be attained~\cite{kroutvar04}. In this scenario, provided ultra-fast spin control is again feasible~\cite{berezovsky08}, a potential figure of merit $\ll 10^{-3}$ could be possible, approaching that necessary for fully scalable fault-tolerant quantum computation~\cite{Nielsen00}.

\section{Summary}

Coherence is an important concept in quantum physics. It underpins many of the uniquely quantum effects we can observe, and is not confined to obvious wave-like phenomena such as Young's double slit experiment. It is crucial for describing solid state nanosystems and leads to some profoundly new applications such as QIP.

\section{Acknowledgments}

BWL acknowledges support from a Royal Society University Research Fellowship.
We would like to thank the QIPIRC (No. GR/S82176/01) for support. AN is supported by the {\sc EPSRC}.
\\\\

\end{document}